\documentstyle[12pt,psfig]{article}

\def\baselinestretch{1.2}
\begin{document}

\begin{titlepage}

\begin{flushright}
UMN-D-01-1 \\
hep-th/0101120
\end{flushright}
\vfil\vfil

\begin{center}

{\Large {\bf Two-Point Stress-Tensor Correlator in ${\cal N}=1
\,\, \mbox{SYM}_{2+1}$ }}

\vfil
J.R. Hiller,$^a$ S. Pinsky,$^b$ and U. Trittmann$^b$
 \\

\vfil

$^a$Department of Physics\\ University of Minnesota Duluth \\
Duluth, MN 55812, USA\\

\vfil
$^b$Department of Physics \\ Ohio State University, Columbus, OH 43210, USA\\

\vfil
\end{center}

\begin{abstract}
\noindent Recent advances in string theory have highlighted the need for
reliable numerical methods to calculate correlators at strong coupling in
supersymmetric theories. We present a calculation of the correlator 
$\langle 0|T^{++}(r) T^{++}(0) |0\rangle $ in
${\cal N} =1 $ SYM theory in 2+1 dimensions. The numerical method we
use is supersymmetric discrete light-cone quantization
(SDLCQ), which preserves the supersymmetry at every order of the approximation
and treats fermions and bosons on the same footing. This calculation is done
at large $N_c$. For small and intermediate $r$ the correlator
converges rapidly for all couplings. At small $r$ the correlator
behaves like $1/r^6$, as expected from conformal field theory.  At large
$r$ the correlator is dominated by the BPS states of the theory. There is, however, a
critical value of the coupling where the large-$r$ correlator goes to zero, suggesting
that the large-$r$ correlator can only be trusted to some finite coupling which
depends on the transverse resolution. We find that this critical coupling grows
linearly with the square root of the transverse momentum resolution. 
\end{abstract}

\vfil\vfil\vfil
\begin{flushleft}
January 2001
\end{flushleft}

\end{titlepage}
\renewcommand{\baselinestretch}{1.05}  

\section{Introduction}
Our original motivation to study correlators of the energy momentum tensor 
\cite{ahlp99,hlpt00} was
the discovery that certain field theories admit concrete realizations as a
string theory on a particular background
\cite{adscft}. Attempts to apply these correspondences to study the details of
these theories have only met with limited success so far. The problem stems from
the fact that this correspondence relates weakly coupled supergravity and 
strongly coupled SYM theory. Unfortunately we only have firm control of either
theory in the weak coupling limit. The objective of our program is to improve
this situation substantially. 

Previously we showed that
Supersymmetric Discrete Light Cone Quantization (SDLCQ) \cite{mss95,hak95}
can be used to solve supersymmetric field
theories in the strong coupling limit \cite{alp98,alp98a,lup99}. This then
allowed us to make a quantitative comparison between the strongly coupled ${\cal
N} = (8,8)$ SYM theory and the supergravity approximation of the string theory
\cite{ahlp99,hlpt00} in 1+1 dimensions.  The SDLCQ approach
works particularly well in 1+1 dimensions; however, it can be extended to more
dimensions. Recently, we solved for the spectrum and wave functions of 
${\cal N}=1$ SYM in 2+1 dimensions \cite{hhlp99,alp99b}.

Aside from our numerical solutions, there has been very little work on
solving SYM theories using methods that might be described as being from first
principles.  While selected properties of these theories
have been investigated, one needs the complete solution of the theory to
calculate the correlators. By a ``complete solution'' we mean the spectrum and the
wave functions of the theory in some well-defined basis. The SYM
theories that are needed for the correspondence with supergravity and string
theory have typically a high degree of supersymmetry and 
therefore a large number of fields.  The number of fields significantly increases
the size of the numerical problem, and, therefore, in this first calculation of
correlators in 2+1 dimensions we consider only ${\cal N}=1$ SYM. 

A convenient quantity that can be computed on both sides of the correspondence
is the correlation function of a gauge invariant operator \cite{GKP,Wit}. We
will focus on two-point functions of the stress-energy tensor.  This turns out
to be a very convenient quantity to compute for reasons that are discussed in
\cite{ahlp99}.  Following the procedure that we used in our calculation in 1+1
dimensions \cite{ahlp99,hlpt00}, we continue the results to Euclidean
space.  The correlator of the energy momentum
operator has been studied in conformal field theory in 2+1 dimensions
\cite{osp93}, and this provides a reference point for our results.
The structure of the correlators in conformal field theory is particularly
simple in the collinear limit $x_\perp\rightarrow 0$, and we therefore find 
it convenient to work in this limit. 
From results in conformal field theory we expect that
correlators behave as $1/r^6$ at small $r$, where we are probing deep inside
the bound states.  We have confirmed this
$1/r^6$ behavior by an analytic calculation of the free-particle correlator
in the DLCQ formalism \cite{BPP}. 

The contributions of individual bound states have a characteristic length scale
corresponding to the size of the bound states.  On dimensional grounds one can
show that the power behavior of the correlators are reduced by one power of $r$; so
for individual bound states the correlator behaves like $1/r^5$ for small $r$. It
then becomes a nontrivial check to see that at small $r$ the contributions of the
bound states add up to give the expected $1/r^6$ behavior.  We find this expected
result as well as the characteristic rapid convergence of SDLCQ at both small and
intermediate values of $r$.

At large $r$ the
correlator is controlled by the massless states of the theory. In this theory
there are two types of massless states. At zero coupling all the states of the $1+1$
dimensional theory are massless, and for non-vanishing coupling 
the massless states of the
$1+1$ theory are promoted to massless states of the $2+1$ dimensional theory
\cite{alp99b}. These states are BPS states and are exactly annihilated by one of
the supercharges. This is perhaps the most interesting part of this calculation
because the BPS masses are protected by the exact
supersymmetry of the numerical approximation and remain exactly zero at all
couplings. Commonly in modern field theory one uses the BPS states to extrapolate
from weak coupling to strong coupling. While the masses of BPS states remain
constant as functions of the coupling, their wave functions certainly do not. The
calculation of the correlator at large $r$ provides a window to the coupling
dependence of the BPS wave functions.  We find, however, that there is a critical
coupling  where the correlator goes to zero, which depends on
the transverse resolution. A detailed study of this
critical coupling shows that it goes to infinity linearly with the square root of the
transverse resolution. Below the critical coupling the correlator converges rapidly
at large $r$. One possible explanation 
is that this singular behavior signals the breakdown of
the SDLCQ calculation for the BPS wave function at couplings larger than the critical
coupling. If this is correct, calculation of the BPS wave function at stronger
couplings  requires higher transverse resolutions. We note that above the critical
coupling (see Fig.~\ref{k5+6large} below)  we do find convergence of the correlator
at large $r$ but at a significantly slower rate.  

This paper is organized as follows. In section 2 we discuss light cone
quantization and SDLCQ.  The correlators are discussed in section 3 for the
free theory and in section 4 for the full theory.  In section
5 we discuss our numerical results.  A brief conclusion is given in
section 6.

\section{Light-Cone Quantization and SDLCQ}
\label{formulation}

The technique of DLCQ is reviewed in \cite{BPP},
so we will be brief here.  The basic idea of light-cone quantization
is to parameterize space-time using light-cone coordinates $x^+$,
$x^-$, $x^\perp$, and to quantize the theory such that $x^+$ plays 
the role of a time.
In the discrete light-cone approach, we require the momentum $p_- =
p^+$ along the $x^-$ direction to take on discrete values in units of
$P^+/K$, where $P^+$ is the conserved total momentum of the system and
$K$ is an integer usually referred to as the harmonic
resolution \cite{BPP}.
One can think of this discretization as a consequence of compactifying
the $x^-$ coordinate on a circle with a period $2L = {2 \pi K /
P^+}$. Along the direction $x^{\perp}$ 
the transverse momentum is discretized as well;
however, it is treated in a fundamentally different way. 
The transverse resolution is $T$,
and we think of the theory as being compactified on a transverse circle of
length $l$.
Therefore, the transverse momentum is cut off at
$ \pm 2\pi T/l$ and discretized in units of $2\pi/l$. Removal of this transverse
momentum cutoff therefore corresponds to taking the transverse resolution $T$ to
infinity.

The advantage of
discretizing on the light cone is the fact that the dimension of the Hilbert
space becomes finite.  Therefore, the Hamiltonian is a finite
dimensional matrix,
and its dynamics can be solved explicitly.  In SDLCQ one makes the DLCQ
approximation to the supercharges, and these discrete representations
satisfy the
supersymmetry algebra. Therefore SDLCQ enjoys the improved renormalization
properties of supersymmetric theories.  Of course, to recover the continuum
result we must send $K$ and $T$ to infinity and, as luck would have it, we find
that SDLCQ usually converges faster than ordinary DLCQ\@. 
Faster convergence is important because the size of the matrices and,
consequently, the difficulty of the computation grow as the 
resolution is increased.

Let us now review these ideas in the context of a specific
super-Yang-Mills theory.
We start with $2+1$ dimensional ${\cal N}=1$ super-Yang-Mills theory
defined on a
space-time with one transverse dimension compactified on a circle.
The action is
\begin{equation}
S=\int d^2 x \int_0^l dx_\perp \mbox{tr}(-\frac{1}{4}F^{\mu\nu}F_{\mu\nu}+
{\rm i}{\bar\Psi}\gamma^\mu D_\mu\Psi).
\end{equation}
After introducing the light--cone coordinates
$x^\pm=\frac{1}{\sqrt{2}}(x^0\pm x^1)$, decomposing the spinor $\Psi$
in terms of chiral projections
\begin{equation}
\psi=\frac{1+\gamma^5}{2^{1/4}}\Psi,\qquad
\chi=\frac{1-\gamma^5}{2^{1/4}}\Psi
\end{equation}
and choosing the light-cone gauge $A^+=0$, we obtain the action
in the form
\begin{eqnarray}\label{action}
S&=&\int dx^+dx^- \int_0^l dx_\perp
\mbox{tr}\left[\frac{1}{2}(\partial_-A^-)^2+
D_+\phi\partial_-\phi+ {\rm i}\psi D_+\psi+ \right.\nonumber \\
& &
\left.
       \hspace{15mm} +{\rm i}\chi\partial_-\chi+\frac{{\rm i}}{\sqrt{2}}\psi
D_\perp\phi+
\frac{{\rm i}}{\sqrt{2}}\phi D_\perp\psi \right].
\end{eqnarray}
A simplification of the
light-cone gauge is that the
non-dynamical fields $A^-$ and $\chi$ may be explicitly
solved from their Euler--Lagrange equations of motion
\begin{equation}
A^-=\frac{g_{\rm YM}}{\partial_-^2}J=
\frac{g_{\rm YM}}{\partial_-^2}\left(i[\phi,\partial_-\phi]+2\psi\psi\right), \quad
\chi=-\frac{1}{\sqrt{2}\partial_-}D_\perp\psi.\nonumber
\end{equation}

These expressions may be used to express any operator
in terms of the physical degrees of freedom only.
In particular, the light-cone energy, $P^-$, and momentum
operators, $P^+$,$P^{\perp}$,
corresponding to  translation
invariance in each of the coordinates
$x^\pm$ and $x_\perp$ may be calculated explicitly as
\begin{eqnarray}\label{moment}
P^+&=&\int dx^-\int_0^l dx_\perp\mbox{tr}\left[(\partial_-\phi)^2+
{\rm i}\psi\partial_-\psi\right],\\
P^-&=&\int dx^-\int_0^l dx_\perp\mbox{tr}
\left[-\frac{g_{\rm YM}^2}{2}J\frac{1}{\partial_-^2}J-
          \frac{{\rm i}}{2}D_\perp\psi\frac{1}{\partial_-}D_\perp\psi\right],\\
P_\perp &=&\int dx^-\int_0^l
dx_\perp\mbox{tr}\left[\partial_-\phi\partial_\perp\phi+
          {\rm i}\psi\partial_\perp\psi\right].
\end{eqnarray}
The light-cone supercharge in this theory
is a two-component Majorana spinor, and may be conveniently
decomposed in terms of its chiral projections
\begin{eqnarray}\label{sucharge}
Q^+&=&2^{1/4}\int dx^-\int_0^l
dx_\perp\mbox{tr}\left[\phi\partial_-\psi-\psi\partial_-
                 \phi\right],\\
\label{sucharge-}
Q^-&=&2^{3/4}\int dx^-\int_0^l dx_\perp\mbox{tr}\left[2\partial_\perp\phi\psi+
          g_{\rm YM}\left({\rm
i}[\phi,\partial_-\phi]+2\psi\psi\right)\frac{1}{\partial_-}\psi\right].
\end{eqnarray}
The action (\ref{action}) gives the following canonical
(anti-)commutation relations for
propagating fields for large $N_c$ at equal $x^+$:
\begin{equation}
\left[\phi_{ij}(x^-,x_\perp),\partial_-\phi_{kl}(y^-,y_\perp)\right]=
\left\{\psi_{ij}(x^-,x_\perp),\psi_{kl}(y^-,y_\perp)\right\}=
\frac{1}{2}\delta(x^- -y^-)\delta(x_\perp -y_\perp)\delta_{il}\delta_{jk}.
\label{comm}
\end{equation}
Using these relations one can check the supersymmetry algebra
\begin{equation}
\{Q^+,Q^+\}=2\sqrt{2}P^+,\qquad \{Q^-,Q^-\}=2\sqrt{2}P^-,\qquad
\{Q^+,Q^-\}=-4P_\perp.
\label{superr}
\end{equation}

In solving for mass eigenstates, we will consider 
only states which have vanishing transverse momentum,
which is possible since the total transverse momentum operator
is kinemat\-ical.\footnote{Strictly speaking, on a transverse
cylinder, there are separate sectors with total
transverse momenta $2\pi N_\perp/L$; we consider only one of them, $N_\perp=0$.}
On such states, the light-cone supercharges
$Q^+$ and $Q^-$ anti-commute with each other, and the supersymmetry algebra
is equivalent to the ${\cal N}=(1,1)$ supersymmetry
of the dimensionally reduced ({\em i.e.}, two-dimensional) theory \cite{mss95}.
Moreover, in the $P_{\perp} = 0$ sector,
the mass squared operator $M^2$ is given by
$M^2=2P^+P^-$.

As we mentioned earlier, in order to render the bound-state equations
numerically tract\-able, the transverse
momenta of partons must be truncated.
First, we introduce the Fourier expansion for the fields $\phi$ and $\psi$,
where the transverse space-time coordinate $x_{\perp}$ is periodically
identified
\begin{eqnarray}
\lefteqn{
\phi_{ij}(0,x^-,x_\perp) =} & & \nonumber \\
& &
\frac{1}{\sqrt{2\pi l}}\sum_{n^{\perp} = -\infty}^{\infty}
\int_0^\infty
       \frac{dk^+}{\sqrt{2k^+}}\left[
       a_{ij}(k^+,n^{\perp})e^{-{\rm i}k^+x^- -{\rm i}
\frac{2 \pi n^{\perp}}{l} x_\perp}+
       a^\dagger_{ji}(k^+,n^{\perp})e^{{\rm i}k^+x^- +
{\rm i}\frac{2 \pi n^{\perp}}{l}  x_\perp}\right]\,,
\nonumber\\
\lefteqn{
\psi_{ij}(0,x^-,x_\perp) =} & & \nonumber \\
& & \frac{1}{2\sqrt{\pi l}}\sum_{n^{\perp}=-\infty}^{\infty}\int_0^\infty
       dk^+\left[b_{ij}(k^+,n^{\perp})e^{-{\rm i}k^+x^- -
{\rm i}\frac{2 \pi n^{\perp}}{l} x_\perp}+
       b^\dagger_{ji}(k^+,n^\perp)e^{{\rm i}k^+x^- +{\rm i}
\frac{2 \pi n^{\perp}}{l} x_\perp}\right]\,.
\nonumber
\end{eqnarray}
Substituting these into the (anti-)commutators (\ref{comm}),
one finds
\begin{equation}
\left[a_{ij}(p^+,n_\perp),a^\dagger_{lk}(q^+,m_\perp)\right]=
\left\{b_{ij}(p^+,n_\perp),b^\dagger_{lk}(q^+,m_\perp)\right\}=
\delta(p^+ -q^+)\delta_{n_\perp,m_\perp}\delta_{il}\delta_{jk}.
\end{equation}
The supercharges then take the following form:
\begin{eqnarray}\label{TruncSch}
&&Q^+={\rm i}2^{1/4}\sum_{n^{\perp}\in {\bf Z}}\int_0^\infty dk\sqrt{k}\left[
b_{ij}^\dagger(k,n^\perp) a_{ij}(k,n^\perp)-
a_{ij}^\dagger(k,n^\perp) b_{ij}(k,n^\perp)\right],\\
\label{Qminus}
&&Q^-=\frac{2^{7/4}\pi {\rm i}}{l}\sum_{n^{\perp}\in {\bf Z}}\int_0^\infty dk
\frac{n^{\perp}}{\sqrt{k}}\left[
a_{ij}^\dagger(k,n^\perp) b_{ij}(k,n^\perp)-
b_{ij}^\dagger(k,n^\perp) a_{ij}(k,n^\perp)\right]+\nonumber\\
&&+ {{\rm i} 2^{-1/4} {g_{\rm YM}} \over \sqrt{l\pi}}
\sum_{n^{\perp}_{i} \in {\bf Z}} \int_0^\infty dk_1dk_2dk_3
\delta(k_1+k_2-k_3) \delta_{n^\perp_1+n^\perp_2,n^\perp_3}
\left\{ \frac{}{} \right.\nonumber\\
&&{1 \over 2\sqrt{k_1 k_2}} {k_2-k_1 \over k_3}
[a_{ik}^\dagger(k_1,n^\perp_1) a_{kj}^\dagger(k_2,n^\perp_2)
b_{ij}(k_3,n^\perp_3)
-b_{ij}^\dagger(k_3,n^\perp_3)a_{ik}(k_1,n^\perp_1)
a_{kj}(k_2,n^\perp_2) ]\nonumber\\
&&{1 \over 2\sqrt{k_1 k_3}} {k_1+k_3 \over k_2}
[a_{ik}^\dagger(k_3,n^\perp_3) a_{kj}(k_1,n^\perp_1) b_{ij}(k_2,n^\perp_2)
-a_{ik}^\dagger(k_1,n^\perp_1) b_{kj}^\dagger(k_2,n^\perp_2)
a_{ij}(k_3,n^\perp_3) ]\nonumber\\
&&{1 \over 2\sqrt{k_2 k_3}} {k_2+k_3 \over k_1}
[b_{ik}^\dagger(k_1,n^\perp_1) a_{kj}^\dagger(k_2,n^\perp_2)
a_{ij}(k_3,n^\perp_3)
-a_{ij}^\dagger(k_3,n^\perp_3)b_{ik}(k_1) a_{kj}(k_2,n^\perp_2) ]\nonumber\\
&& ({ 1\over k_1}+{1 \over k_2}-{1\over k_3})
[b_{ik}^\dagger(k_1,n^\perp_1) b_{kj}^\dagger(k_2,n^\perp_2)
b_{ij}(k_3,n^\perp_3)
+b_{ij}^\dagger(k_3,n^\perp_3) b_{ik}(k_1,n^\perp_1) b_{kj}(k_2,n^\perp_2)]
       \left. \frac{}{}\right\}. \nonumber \\
\end{eqnarray}
We now perform the truncation procedure; namely, in all sums over the
transverse momenta $n^{\perp}_{i}$ appearing in the above expressions for the
supercharges, we restrict summation to the following allowed momentum
modes: $n^{\perp}_{i}=0,\pm 1 ... \pm T$.  Note that this prescription is
symmetric, in the sense that there are as many positive modes as there are
negative ones. In this way we  retain parity symmetry in the transverse
direction. The longitudinal momenta $k_i=n_i \pi/L$ are restricted by the
longitudinal resolution according to $K=\sum_i n_i$.

There are three commuting
$Z_2$ symmetries.  One of them is the parity in the transverse direction,
\begin{equation}\label{parity}
P: a_{ij}(k,n^\perp)\rightarrow -a_{ij}(k,-n^\perp),\qquad
      b_{ij}(k,n^\perp)\rightarrow b_{ij}(k,-n^\perp).
\end{equation}
The second symmetry \cite{Kutasov93} is with respect to the operation
\begin{equation}\label{Z2}
S: a_{ij}(k,n^\perp)\rightarrow -a_{ji}(k,n^\perp),\qquad
      b_{ij}(k,n^\perp)\rightarrow -b_{ji}(k,n^\perp).
\end{equation}
Since $P$ and $S$ commute with each other, we need only one additional
symmetry $R=PS$ to close the group.
Since $Q^-$, $P$ and $S$ commute with each other, we can diagonalize them
simultaneously. This allows us to diagonalize the supercharge separately  in
the sectors with fixed $P$ and $S$ parities and thus reduce the size
of matrices. Doing this one finds that the roles of $P$ and $S$ are different.
While all the eigenvalues are usually broken into non-overlapping $S$-odd and
$S$-even sectors \cite{bdk93}, the $P$ symmetry leads to a double
degeneracy of massive states (in addition to the usual boson-fermion degeneracy
due to supersymmetry).

\section{Free particle Correlation Functions}

Let us now return to the problem at hand. We would like to compute a
general expression of the form
\begin{equation}
F(x^+,x^-,x^\perp) = \langle 0| T^{++}(x^+,x^-,x^\perp) 
T^{++}(0,0,0)|0 \rangle.
\end{equation}
Here we will calculate the correlator in the collinear limit, that is,
where $x^\perp = 0$. We know from conformal field theory
\cite{osp93}
calculations that this
will produce a much simpler structure. 

The calculation is done by inserting a
complete set of intermediate states $| \alpha \rangle$,
\begin{equation} \label{eq:F}
F(x^+,x^-,x^\perp=0) = \sum_\alpha \langle 0| T^{++}(x^-,0,x^\perp=0)
| \alpha \rangle e^{-iP^-_\alpha x^+} \langle \alpha | T^{++}(0,0,0)
|0 \rangle.
\end{equation}
with energy eigenvalues $P^-_\alpha$.
In \cite{alpp98} we
found that the momentum operator $T^{++}(x)$ is given by
\begin{equation}
T^{++}(x) =  {\rm tr} \left[ (\partial_- \phi)^2 + {1 \over 2} \left(i
\psi \partial_- \psi  - i  (\partial_- \psi) \psi
\right)\right]=T^{++}_B(x)+T^{++}_F(x)\,.
\end{equation}
In terms of the mode operators, we find
\begin{equation}
T^{++}(x^+,x^-,0) | 0 \rangle = {1 \over 2L l} \sum_{n,m}\,\,
\sum_{n_\perp,m_\perp}
T(n,m) e^{-i(P_n^+ + P_m^+)x^-}
   |0\rangle\,,
\end{equation}
where the boson and fermion contributions are given by
\begin{equation}
\frac{L}{\pi}T^{++}_B(n,m) | 0 \rangle = {\sqrt{n m} \over 2}
    {\rm tr} \left[ a^\dagger_{ij}(n,n_\perp) a^\dagger_{ji} (m,m_\perp)
   \right]| 0\rangle
\end{equation}
and
\begin{equation}
\frac{L}{\pi}T^{++}_F(n,m) | 0 \rangle = {(n-m) \over 4}
{\rm tr}\left[b^\dagger_{ij}(n,n_\perp) b^\dagger_{ji}(m,m_\perp)
\right] | 0 \rangle\,.
\end{equation}
Given each $|\alpha\rangle$, the matrix elements in (\ref{eq:F}) can
then be evaluated, and the sum computed.

First, however, it is instructive to do the calculation where the states $|\alpha\rangle$
are a set of free particles with mass $m$. The boson contribution is
\begin{eqnarray}
F(x^+,x^-,0)_B&=& \sum_{n,m,s,t} \left(\frac{\pi}{4L^2 l}\right)^2
\langle 0|{\rm tr}[a(n,n_\perp) a(m,m_\perp)] {\rm tr}[a^\dagger(s,s_\perp)
a^\dagger(t,t_\perp)] |0\rangle \\ \nonumber
&&\times \sqrt{mnst} \, e^{-iP^-_n x^+ -iP^+_n x^--iP^-_m x^+ -iP^+_m x^-}\,,
\end{eqnarray}
where the sum over $n$ implies sums over both $n$ and $n_\perp$, and
\begin{equation}
P^-_n=\frac{m^2 +\left(2 n_\perp \pi/l \right)^2}{2 n \pi/L}
\quad  {\rm and} \quad P_n^+=\frac{n \pi}{L}\,.
\end{equation}
The sums can be converted to integrals which can be explicitly evaluated, and we
find
\begin{equation}
F(x^+,x^-,0)_B=\frac{i}{2(2\pi)^3} m^5 \left(\frac{x^+}{x^-} \right)^2
\frac{1}{x} K^2_{5/2}(mx)\,,
\end{equation}
where $x^2 =2x^-x^+$.
Similarly for the fermions we find
\begin{eqnarray}
F(x^+,x^-,0)_F&=& \sum_{n,m,s,t} \left(\frac{\pi}{8L^2 l}\right)^2
\langle 0|{\rm tr}[b(n,n_\perp) b(m,m_\perp)] {\rm tr}[b^\dagger(s,s_\perp)
b^\dagger(t,t_\perp)] |0\rangle \\ \nonumber
&&\times (m-n) (s-t) \, e^{-iP^-_n x^+ -iP^+_n x^--iP^-_m x^+ -iP^+_m x^-}\,.
\end{eqnarray}
After doing the integrals we obtain
\begin{equation}
F(x^+,x^-,0)_F=\frac{i}{4(2\pi)^3} m^5 \left(\frac{x^+}{x^-} \right)^2
\frac{1}{x}\left[ K_{7/2}(mx)K_{3/2}(mx)-K^2_{5/2}(mx) \right]\,.
\end{equation}
We can continue to Euclidean
space by taking $r = \sqrt{2 x^+ x^-}$ to be real, and, finally, in the 
small-$r$ limit we find
\begin{equation}
\left(\frac{x^-}{x^+} \right)^2 F(x^+,x^-,0)=\frac{-3 i}{8 (2\pi)^2}
\frac{1}{r^6}\,,
\end{equation}
which exhibits the expected $1/r^6$ behavior.

\section{Correlation Function in SDLCQ}

Now  let us return to the calculation using the bound-state solution
obtained from SDLCQ. It is convenient to write
\begin{equation} \label{eq:Fdiscrete}
F(x^+,x^-,0)=\sum_{n,m,s,t} \,\,\left(\frac{\pi}{2L^2 l}\right)^2
\langle 0|\frac{L}{\pi}T(n,m) e^{-iP^-_{op}x^+-iP^+x^-}
            \frac{L}{\pi} T(s,t)| 0 \rangle\,,
\end{equation}
where $P^-_{op}$ is the Hamiltonian operator.
We again insert a complete set of bound states $|\alpha\rangle $
with light-cone energies $P_\alpha^-=(M_\alpha^2+P_\perp^2)/P^+$
at resolution K (and therefore $P^+=\pi K/L$) and with total 
transverse momentum $P_\perp=2N_\perp\pi/l$. We also define
\begin{equation}
|u\rangle = N_u \frac{L}{\pi}
  \sum_{n,m}\delta_{n+m,K}\delta_{n_\perp+m_\perp,N_\perp}T(n,m) |0 \rangle\,,
\end{equation}
where $N_u$ is a normalization factor such that $\langle u|u \rangle=1$. 
It is straightforward to calculate the normalization, and we find
\begin{equation}
\frac{1}{N_u^2}=\frac{K^3}{8} (1-\frac{1}{K}) (2T+1)\,.
\end{equation}
The correlator (\ref{eq:Fdiscrete}) becomes
\begin{equation}
F(x^+,x^-,0)=\sum_{K,N_\perp,\alpha} \left(\frac{\pi}{2L^2 l}\right)^2 e^{-iP^-_\alpha
x^+-iP^+x^-} \frac{1}{N_u^2} |\langle u|\alpha\rangle |^2\,.
\end{equation}

We will calculate the matrix element $\langle u|\alpha \rangle$ at fixed
longitudinal resolution $K$ and transverse momentum $N_\perp=0$. 
Because of transverse boost invariance
the matrix element does not contain any explicit dependence on $N_\perp$. 
To leading order in
$1/K$ the explicit dependence of the matrix element on $K$ is $K^3$;
it also contains a factor of $l$, the transverse length scale
To separate these dependencies, we write $F$ as
\begin{equation}
F(x^+,x^-,0)=\frac{1}{2 \pi} \sum_{K,N_\perp,\alpha} \frac{1}{2L} 
\frac{1}{l} \left(\frac{\pi K}{L}\right)^3
  e^{-iP^-_\alpha x^+ -iP^+ x^-}
  \frac{|\langle u|\alpha\rangle|^2}{l K^3 |N_u|^2} \,.
\end{equation}

We can now do the sums over $K$ and $N_\perp$ as integrals over the 
longitudinal and transverse momentum components
$P^+=\pi K/L$ and $P^\perp= 2 \pi N_\perp/l$.  We obtain
\begin{equation}
\frac{1}{\sqrt{-i} }\left(\frac{x^-}{x^+} \right)^2 F(x^+,x^-,0)
=\sum_\alpha 
\frac{1}{2 (2\pi)^{5/2}}\frac{M_\alpha^{9/2}}{\sqrt{r}}K_{9/2}(M_\alpha r)
\frac{|\langle u|\alpha\rangle|^2}{lK^3 |N_u|^2} 
\label{master}
\end{equation}
In practice, the full sum over $\alpha$ is approximated by a Lanczos
\cite{Lanczos} iteration technique \cite{hlpt00} that eliminates
the need for full diagonalization of the Hamiltonian matrix.
For the present case, the number of iterations required was on the
order of 1000.

Looking back at the calculation for the free particle, we see that there are
two independent sums over transverse momentum, after the contractions are
performed. One would expect that the transverse dimension is
controlled by the dimensional scale of the bound state ($R_B$) and therefore
the correlation should scale like $1/ r^4 R_B^2$. However, because of
transverse boost invariance, the matrix element must be independent of the
difference of the transverse momenta and therefore must scale as
$1/r^5R_B$.

\section{Numerical Results}

\begin{figure}
\begin{tabular}{cc}
\psfig{file=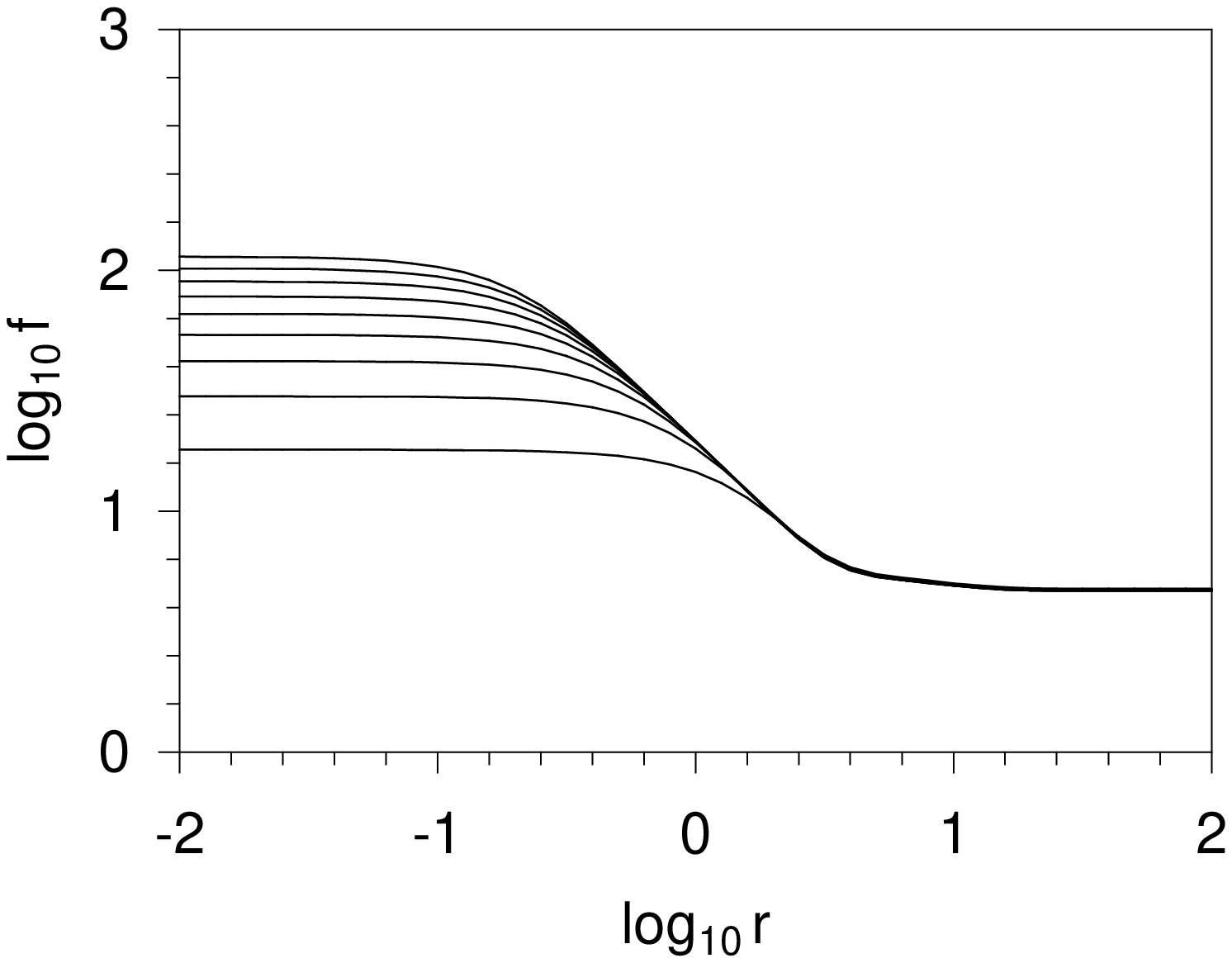,width=3.2in}  &
\psfig{file=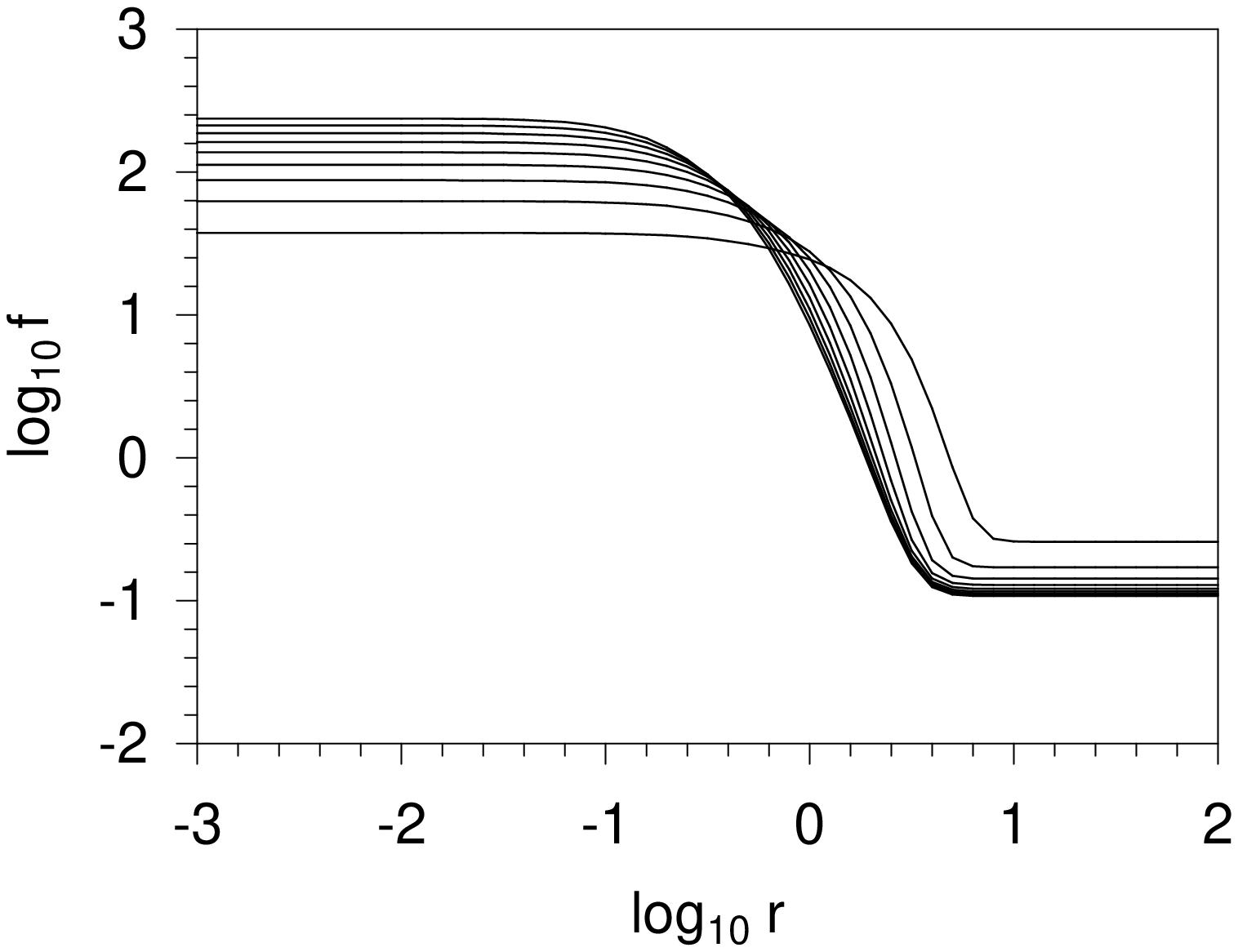,width=3.2in}  \\
(a) & (b)
\end{tabular}
\caption{The log-log
plot of the correlation function $f\equiv r^5\langle T^{++}(x) T^{++}(0) \rangle
\left({x^- \over x^+} \right)^2 \frac{16\pi^3}{105}{{K^3 l} \over \sqrt{-i}}$
vs.\ $r$ (a) in units where $g=g_{\rm YM} \sqrt{N_c l}/2\pi^{3/2} = 0.10$ for $K=4$
and
$T=1$ to 9; (b) in units where $g=g_{\rm YM} \sqrt{N_c l}/2\pi^{3/2} = 1$ for $K=5$
and
$T=1$ to 9.
\label{smallr}}
\end{figure}

\begin{figure}
\centerline{
\psfig{file=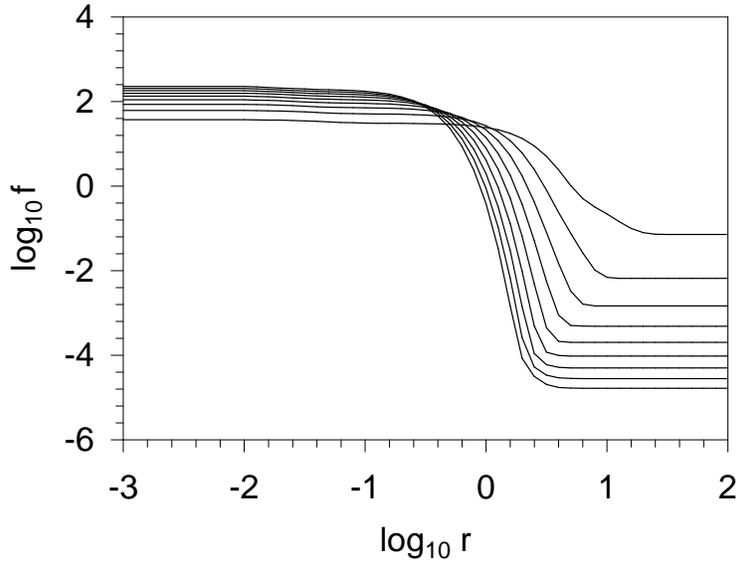,width=4in}}
\caption{Same as Fig.~\ref{smallr}(b), but for 
$g=g_{\rm YM} \sqrt{N_c l}/2\pi^{3/2}= 10$.
\label{k5g1000}}
\end{figure}

The first important numerical test is the small-$r$ behavior of the correlator.
Physically we expect that at small $r$ the bound states should behave as free
particles, and therefore the correlator should have the behavior of the free
particle correlator which goes like $1/r^6$. We see in (\ref{master}) that the
contributions of each of the bound states behaves like $1/r^5$.
Therefore, to get the $1/r^6$ behavior of the free theory, the bound
states must work in concert at small $r$. It
is clear that this cannot work all the way down to $r=0$ 
in the numerical calculation. At very small $r$
the most massive state allowed by the numerical approximation will
dominate, and the correlator must behave like
$1/r^5$. To see what happens at slightly larger $r$ it is useful to
consider the behavior at small coupling. There, the larger masses go
like
\begin{equation}
M_\alpha \simeq \sum_i \frac{(k^\perp_i)^2}{2P^+}\,.
\end{equation}
Consequently, as we remove the $k^\perp$ cutoff, {\em i.e.}~increase the 
transverse
resolution $T$, more and more massive bound states will contribute, and the
dominant one will take over at smaller and smaller $r$ leading to the
expected $1/r^6$. This is exactly what we see happening in 
Fig.~\ref{smallr} at weak coupling with longitudinal resolution $K=4$ and 5.
%
%
The correlator converges from below at small $r$ with increasing $T$, and
in the region $ -0.5 \leq \log r \leq 0.5 $ the plot of $r^5$ times the
correlator falls like $ 1/r$. In Fig.~\ref{k5g1000} at resolution $K=5$ we see
the same behavior for strong coupling ($g =g_{\rm YM} \sqrt{N_c l}/2\pi^{3/2} =10$)
but now at smaller $r$ ($\log r \simeq -0.5 $) as one would expect. 
%
%
Again at strong
coupling we see that the correlator converges quickly and from
below in $T$.  All indications are that at small $r$ the correlators are well
approximated by SDLCQ, converge rapidly, and show the behavior that one
would expect on general physical grounds. This gives us confidence to go
on to investigate the behavior at large $r$.

The behavior for large $r$ is governed by the massless states. From earlier
work \cite{hhlp99,alp99b} on the spectrum of this theory we know that
there are two types of massless states.  At $g=0$ the
massless states are  a reflection of all the states of the dimensionally
reduced theory in $1+1$. In 2+1 dimensions these states behave as $g^2
M^2_{1+1}$. We expect therefore that for $g \simeq 0$ there should be no
dependence of the correlator on the transverse momentum cutoff
$T$ at large $r$. In Fig.~\ref{smallr}(a) this behavior is clearly evident. 

At all couplings there are exactly massless states which are the BPS states of
this theory, which has zero central charge. These states are destroyed by one
supercharge, $Q^-$, and not the other, $Q^+$.  From earlier
work \cite{hhlp99} on the spectrum we saw that the number of BPS states is
independent of the transverse resolution and equal to $2K-1$. Since these
states are exactly massless at all resolutions, transverse and longitudinal
convergence of these states cannot be investigated using the spectrum. 
These states do have a complicated dependence on the coupling $g$ through 
their wave function, however. This is a
feature so far not encountered in DLCQ \cite{BPP}.   
In previous DLCQ calculations one always
looked to the convergence of the spectrum as a measure of the convergence of the
numerical calculation. Here we see that it is the correlator at large $r$
that provides a window to study the convergence of the  wave functions of the BPS
states. In Fig.~\ref{k5g1000} we see that the correlator converges
from above at large $r$ as we increase $T$. 

We also note that the correlator at large $r$ is significantly smaller than at
small $r$, particularly at strong coupling. In our initial study of the BPS states
\cite{alp99b} we found that at strong coupling the average number of particles
in these BPS states is large. Therefore the two particle
components, which are the only components the $T^{++}$ correlator sees, are
small.

\begin{figure}
\begin{tabular}{cc}
\psfig{file=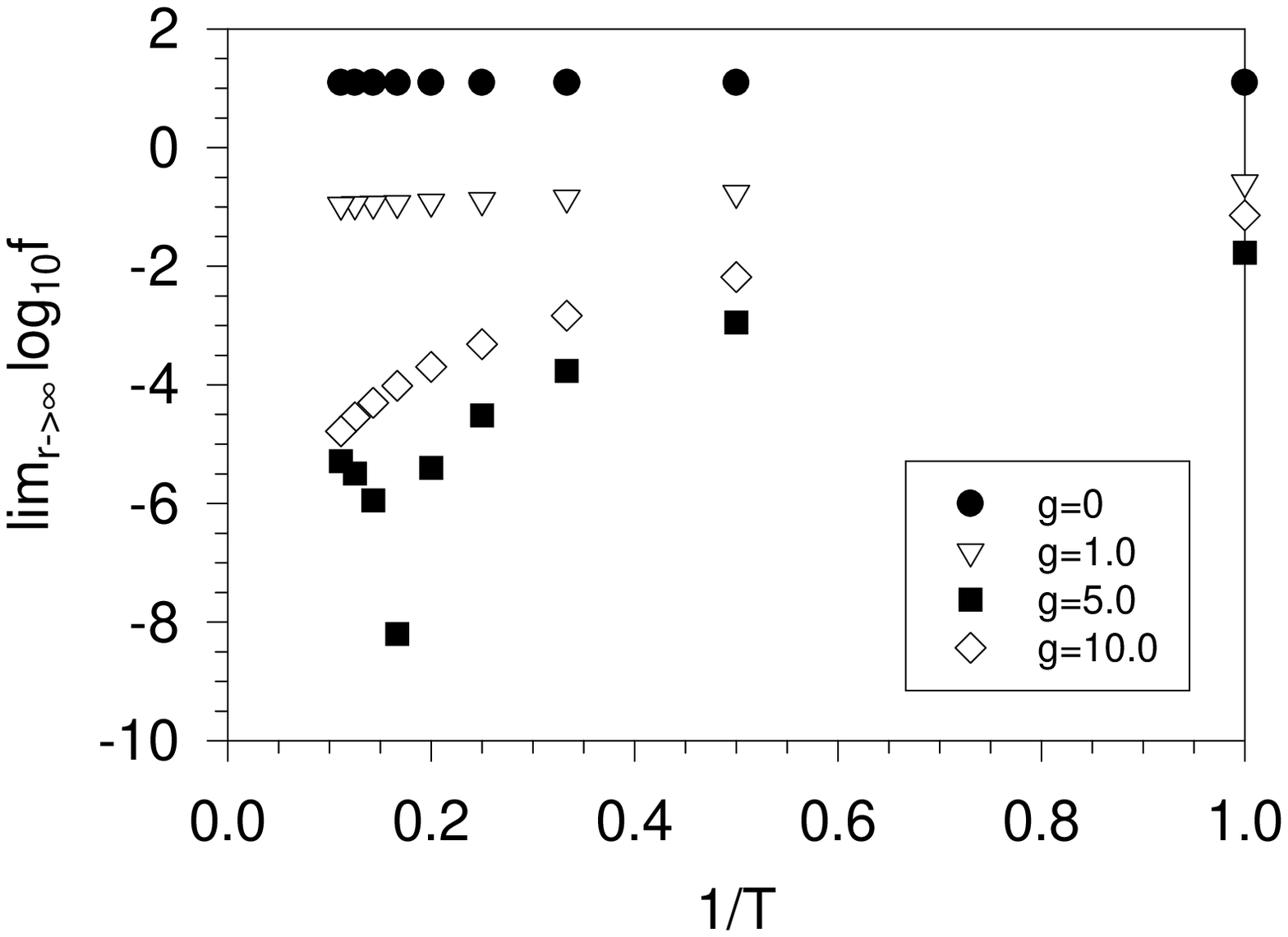,width=3.2in}  &
\psfig{file=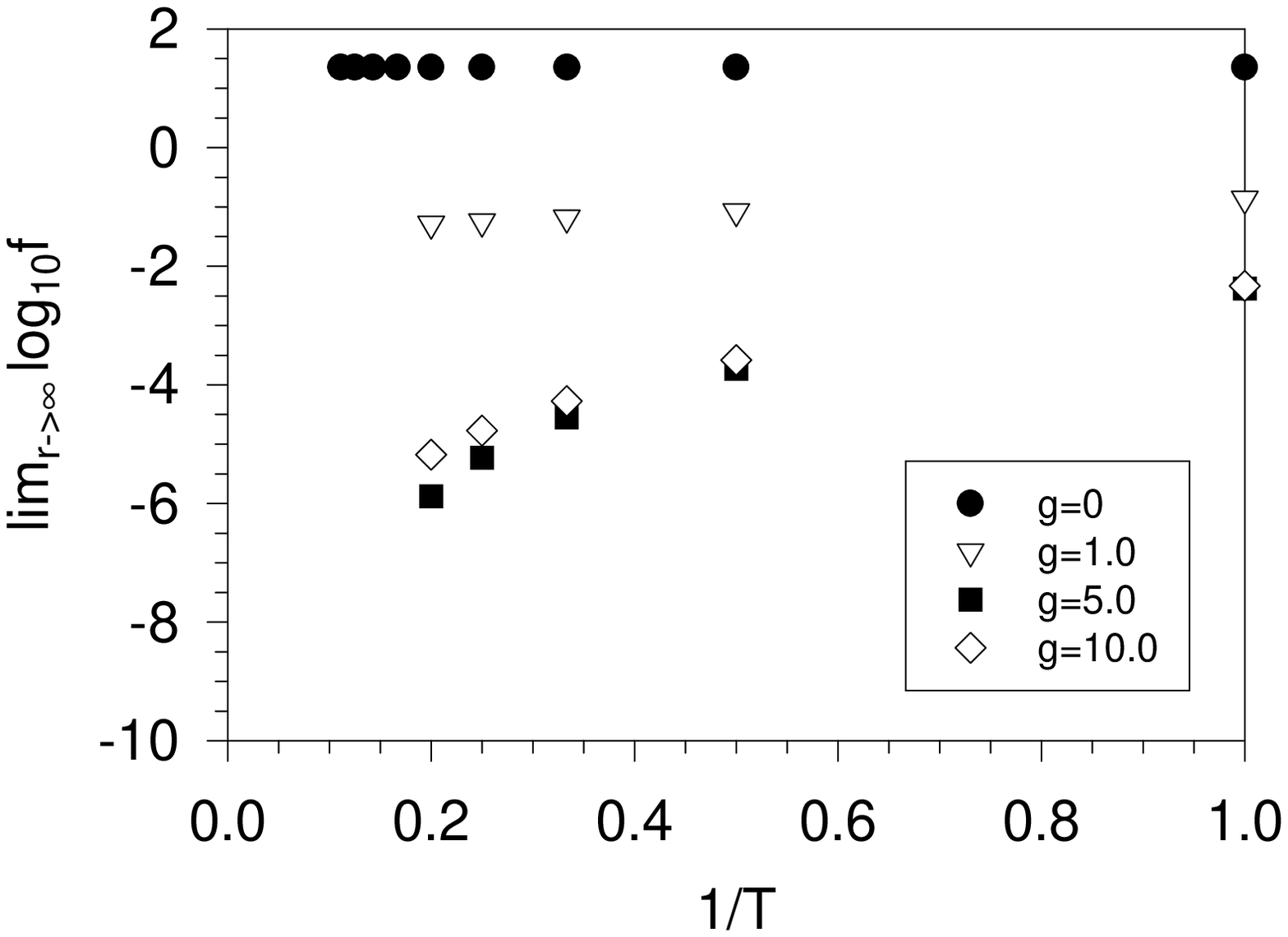,width=3.2in}  \\
(a) & (b)
\end{tabular}
\caption{The large-$r$ limit of the log of 
the correlation function $f\equiv r^5\langle T^{++}(x) T^{++}(0) \rangle
\left({x^- \over x^+} \right)^2 \frac{16\pi^3}{105}{{K^3 l} \over \sqrt{-i}}$
vs.\ $1/T$ for (a) $K=5$ and (b) $K=6$ and for various values of the
coupling $g=g_{\rm YM} \sqrt{N_c l}/2\pi^{3/2}$.
\label{k5+6large}}
\end{figure}
\begin{figure}
\begin{tabular}{cc}
\psfig{file=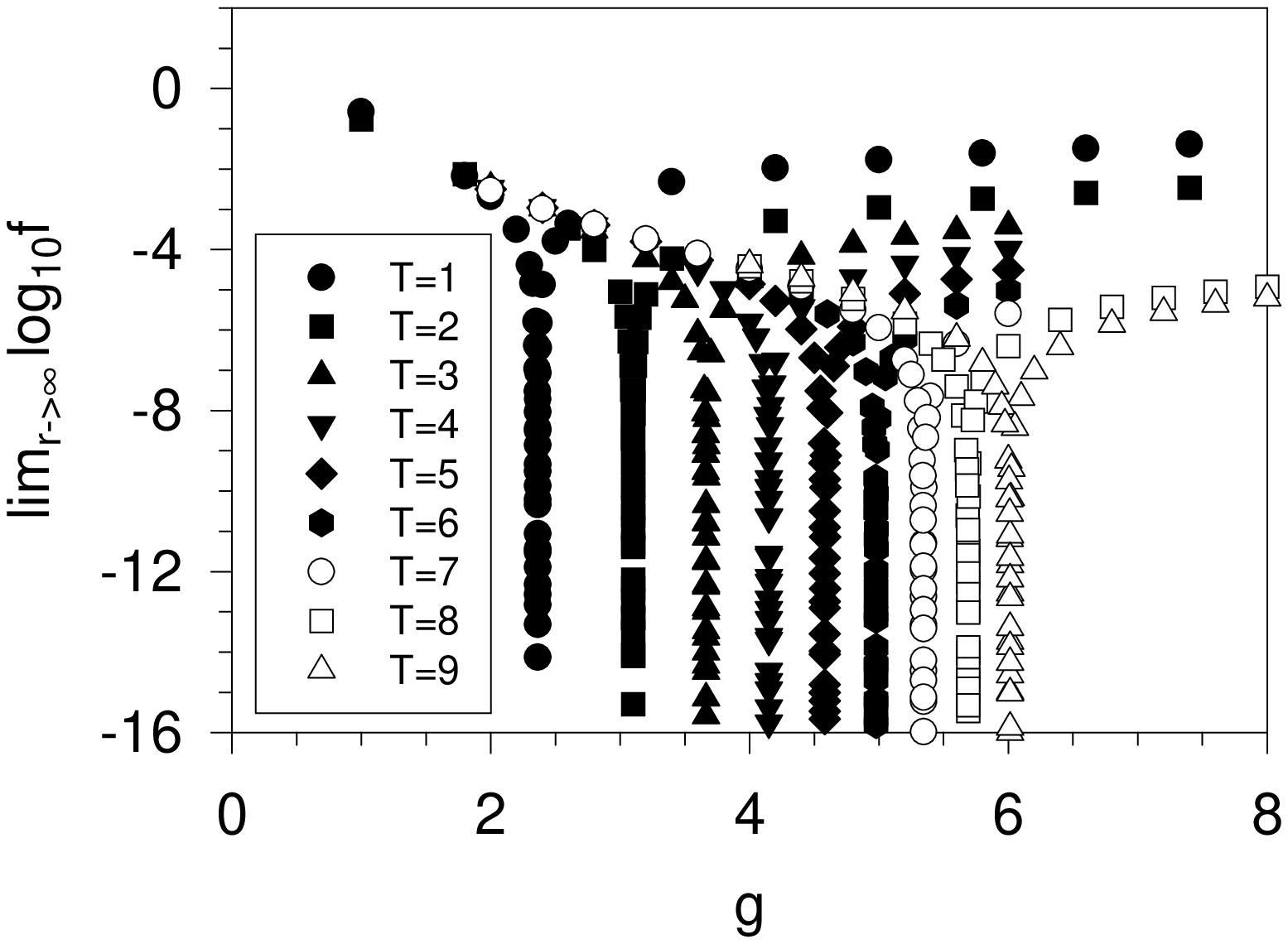,width=3.2in}  &
\psfig{file=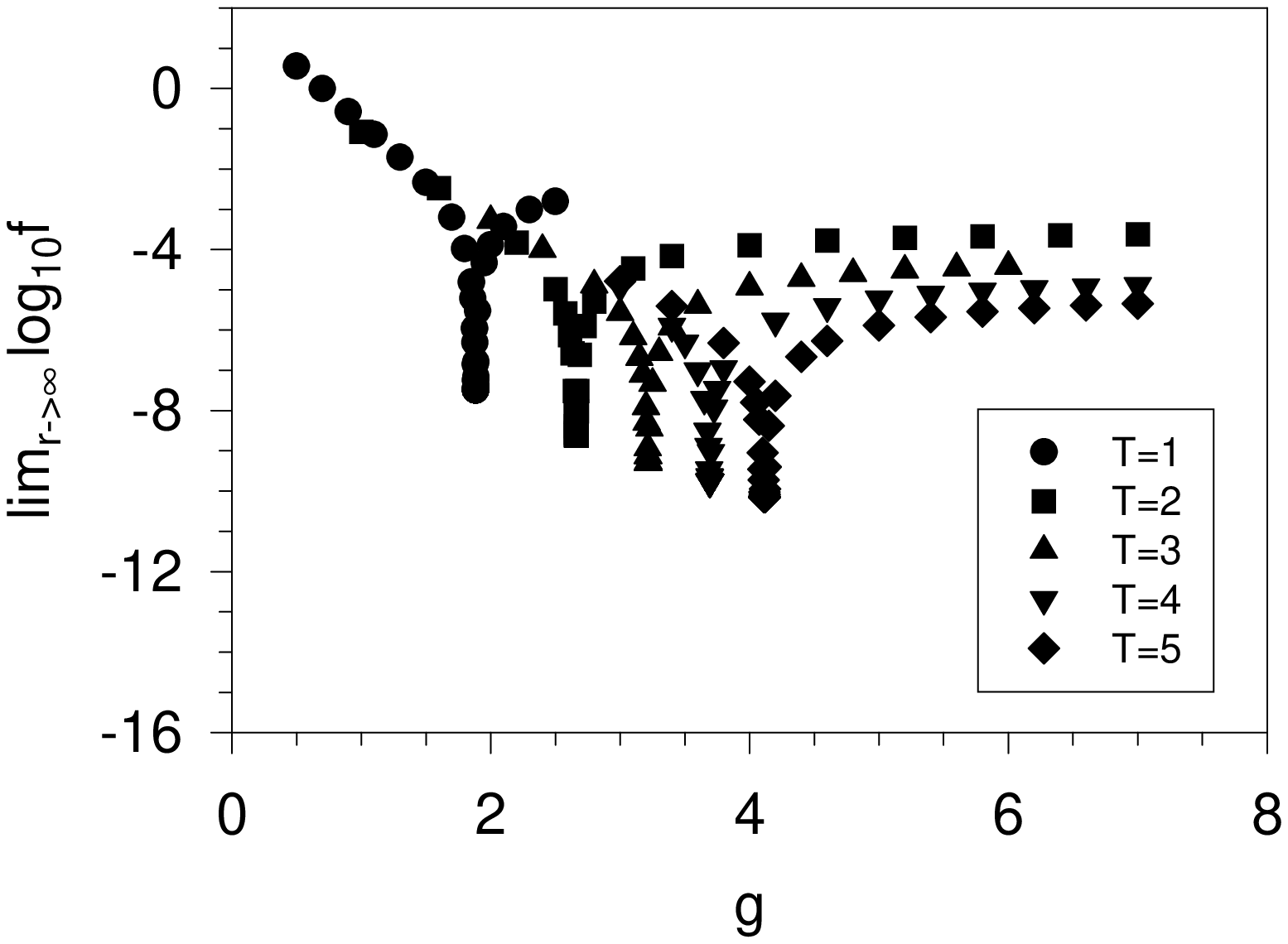,width=3.2in}  \\
(a) & (b)
\end{tabular}
\caption{The large-$r$ limit of
the correlation function $f\equiv r^5\langle T^{++}(x) T^{++}(0) \rangle
\left({x^- \over x^+} \right)^2 \frac{16\pi^3}{105}{{K^3 l} \over \sqrt{-i}}$ 
vs.\ $g=g_{\rm YM} \sqrt{N_c l}/2\pi^{3/2}$ for (a) $K=5$ (b) $K=6$ and for various 
values of the transverse resolution $T$.
\label{k5+6crit}}
\end{figure}

The coupling dependence of the large-$r$ limit of the correlator is much more 
interesting than we would have expected based on our previous work on the 
spectrum. To see this
behavior we study the large-$r$ behavior of the correlator at fixed $g$ as a
function of the transverse resolution $T$ and at fixed $T$ as a function of the
coupling $g$. We see a hint that something unusual is occurring in
Fig.~\ref{k5+6large}. For values of the coupling up to about $g=1$ we see the
typical rapid convergence in the transverse momentum cutoff; however, 
at larger coupling
the convergence appears to deteriorate, and we see that for
$g=5$ the correlator is smaller than at $g=10$. We see this same behavior
at both $K=5$ and $K=6$.\footnote {We do not see this behavior at
$K=4$, but it is not unusual for effects to appear only at a large enough 
resolution in SDLCQ.} 
In Fig.~\ref{k5+6crit} we see that the correlator does not in fact
decrease monotonically with $g$ but rather has a singularity at a
particular value of the coupling which is a function of $K$ and $T$.  
Beyond the singularity the correlator again appears to behave well. 

\begin{figure}
\begin{tabular}{cc}
\psfig{file=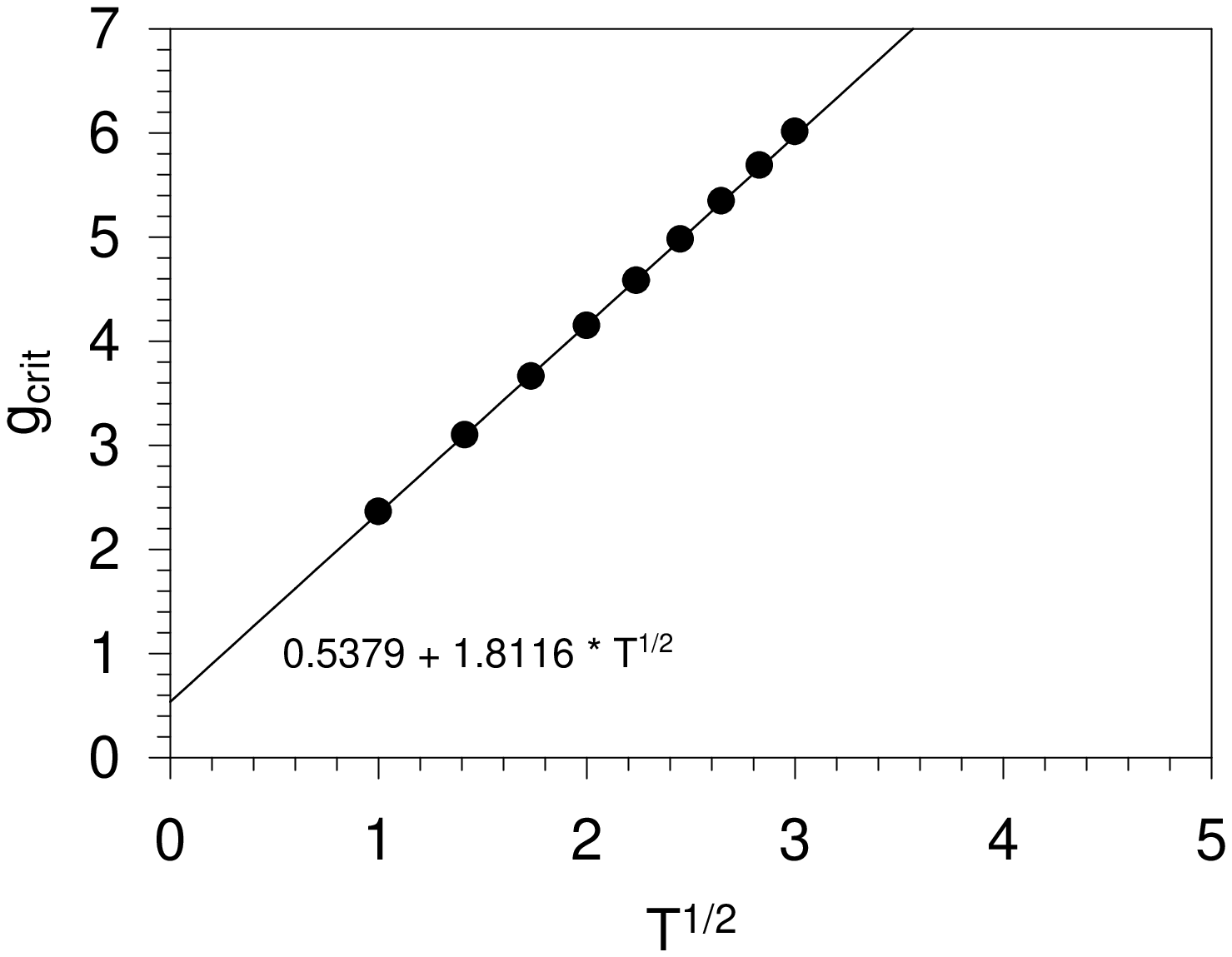,width=3in}  &
\psfig{file=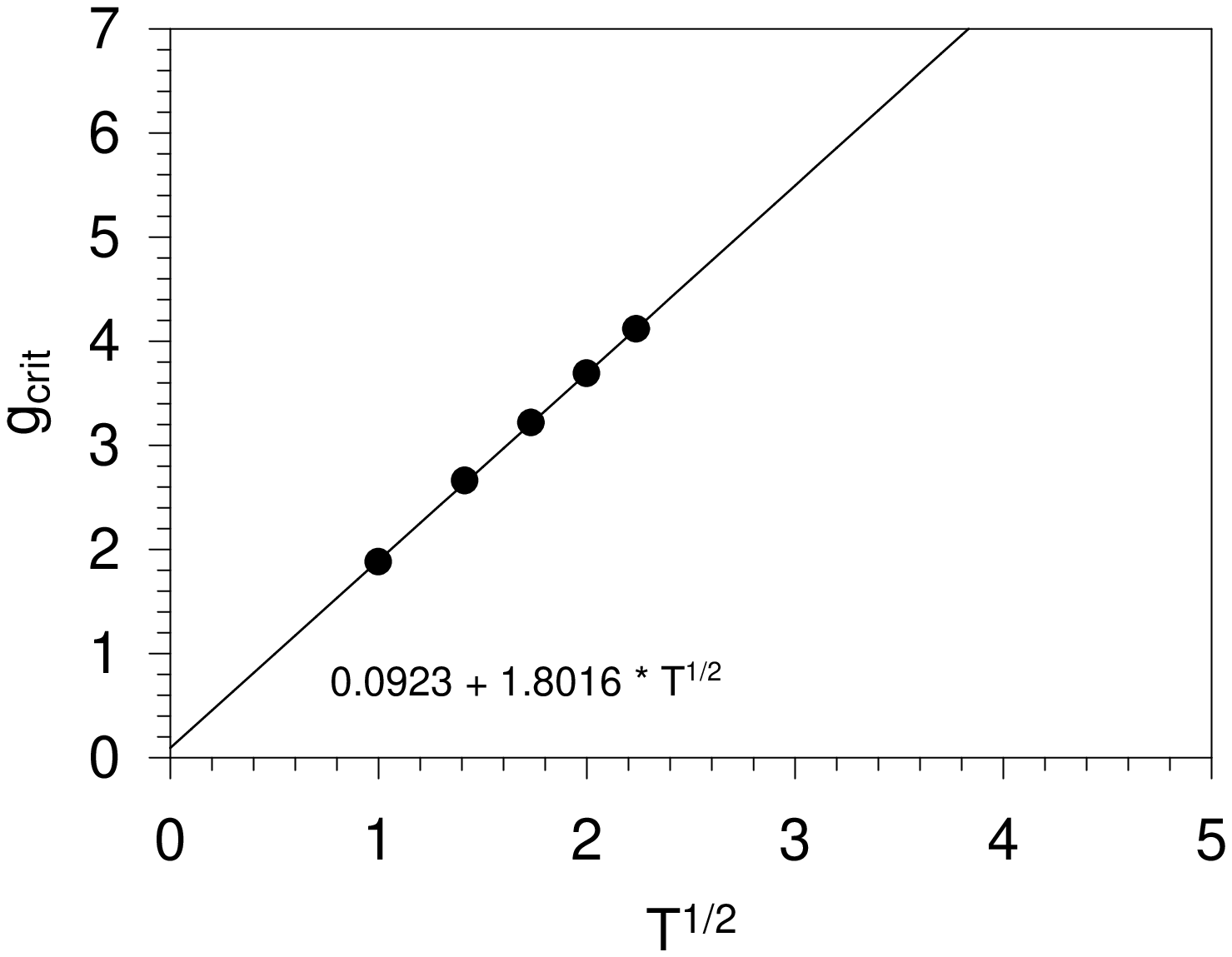,width=3in}  \\
(a) & (b)
\end{tabular}
\caption{
Critical coupling $g_{\rm crit}$ versus $\sqrt{T}$ for (a) $K=5$ and (b) $K=6$.
\label{k5+6critfit2}}
\end{figure}

If we plot the `critical' couplings, at which the correlator goes to zero,
versus $\sqrt{T}$, as in Fig.~\ref{k5+6critfit2}, 
we see that they lie on a straight 
line, {\em i.e.}~this coupling is a linear function
of $\sqrt{T}$ in both cases, $K=5$ and 6. 
Consequently, the `critical' coupling goes to infinity in the
transverse continuum limit.
It appears as though we have encountered a finite transverse cutoff effect. The most
likely conclusion is that our numerical calculation of the BPS wave function is only
valid for $g < g_{\rm crit}(T)$. While the large-$r$ correlator does converge above
the critical coupling, it is unclear at this time if it has any significance. It
might have been expected that one would need larger and larger transverse resolution
to probe the strong coupling region, the occurrence of the singular behavior that we
see is a surprise, and we have no detailed explanation for it at this time. We see
no evidence of a singular behavior at small or intermediate $r$. This indicates, but
does not prove, that our calculations of the massive bound states is valid at all
$g$.

We do not seem to see a region dominated the massive bound states, that is, a
region where $r$ is large enough that we see the structure of the
bound states but small enough that the correlator is not dominated by the massless
states of the theory. Such a region might give us other important information
about this theory.

\section{Conclusion}

In this work we calculate the correlator $\langle 0|T^{++}(x) T^{++}(0)|0
\rangle$ in ${\cal N}=1, $ SYM in 2+1 dimension at large $N_c$ in the collinear
limit.  We find that the free-particle correlator behaves like $1/r^6$, in
agreement with results from conformal field theory.
The contribution from an individual bound state is found to
behave like $1/r^5$, and at small $r$ such contributions conspire to reproduce
the conformal field theory result $1/r^6$. We do not seem to 
find an intermediate region in
$r$ where the correlator behaves as $1/r^5$, reflecting the behavior of the
individual massive bound states. 

At large
$r$ the correlator is dominated by the massless BPS states of the theory. We find
that as a function of $g$ the large-$r$ correlator has a critical value 
of $g$ where it abruptly drops to zero. 
We have investigated this singular behavior and find that at fixed longitudinal
resolution the critical coupling grows linearly with $\sqrt{T}$. We conjecture that
this critical coupling signals the breakdown of SDLCQ at sufficiently strong coupling
at fixed transverse resolution, $T$. While this might not be surprising in general,
it is surprising that the behavior appears in the BPS wave functions and that we see
no sign of this behavior in the massive states. We find that above the critical
coupling the correlator still converges but significantly slower. It is unclear at
this time if we should attach any significance to the correlator in this region. 

This calculation emphasizes the importance of BPS wave functions which carry important
coupling dependence, even though the mass eigenvalues are independent of the
coupling. We will discuss the spectrum, the wavefunctions and associated
properties of all the low energy bound states of
${\cal N}=1$ SYM  in 2+1 dimensions in a subsequent paper \cite{hpt01}.

A number of computational improvements have been implemented in our code to allow
us to make these detailed calculations. The code now fully utilizes the 
three known
discrete symmetries of the theory, namely supersymmetry, transverse parity $P$, 
Eq.~(\ref{parity}), and the $Z_2$ symmetry $S$,
Eq.~(\ref{Z2}).
This reduces the dimension of the Hamiltonian matrix by a factor of 8. 
Other, more efficient storage techniques allow us
to handle on the order of 2,000,000 states in this calculation, which has been
performed on a single processor Linux workstation. Our improved storage techniques
should allow us to expand this calculation to include higher supersymmetries
without a significant expansion of the code or computational power. We
remain hopeful that porting to a parallel machine will allow us to tackle problems
in full 3+1 dimensions.   

\section*{Acknowledgments}
This work was supported in part by
the US Department of Energy.

\end{document}